\documentclass[sn-mathphys,Numbered]{sn-jnl}

\usepackage{graphicx}%
\usepackage{multirow}%
\usepackage{amsmath,amssymb,amsfonts}%
\usepackage{amsthm}%
\usepackage{mathrsfs}%
\usepackage[title]{appendix}%
\usepackage{xcolor}%
\usepackage{textcomp}%
\usepackage{manyfoot}%
\usepackage{booktabs}%
\usepackage{algorithm}%
\usepackage{algorithmicx}%
\usepackage{algpseudocode}%
\usepackage{listings}%
\usepackage{booktabs}

\begin{document}

\title[Article Title]{Can AI Serve as a Substitute for Human Subjects in Software Engineering Research?}

\author*[1]{\fnm{Marco} \sur{Gerosa}}\email{Marco.Gerosa@nau.edu}
\author[2]{\fnm{Bianca} \sur{Trinkenreich}}\email{Bianca.Trinkenreich@oregonstate.edu}
\author[1]{\fnm{Igor} \sur{Steinmacher}}\email{Igor.Steinmacher@nau.edu}
\author[2]{\fnm{Anita} \sur{Sarma}}\email{Anita.Sarma@oregonstate.edu}

\affil*[1]{\orgname{Northern Arizona University}, \country{USA}}
\affil[2]{\orgname{Oregon State University}, \country{USA}}


\abstract{
Research within sociotechnical domains, such as Software Engineering, fundamentally requires a thorough consideration of the human perspective. However, traditional qualitative data collection methods suffer from challenges related to scale, labor intensity, and the increasing difficulty of participant recruitment. This vision paper proposes a novel approach to qualitative data collection in software engineering research by harnessing the capabilities of artificial intelligence (AI), especially large language models (LLMs) like ChatGPT. We explore the potential of AI-generated synthetic text as an alternative source of qualitative data, by discussing how LLMs can replicate human responses and behaviors in research settings. We examine the application of AI in automating data collection across various methodologies, including persona-based prompting for interviews, multi-persona dialogue for focus groups, and mega-persona responses for surveys. Additionally, we discuss the prospective development of new foundation models aimed at emulating human behavior in observational studies and user evaluations. By simulating human interaction and feedback, these AI models could offer scalable and efficient means of data generation, while providing insights into human attitudes, experiences, and performance. We discuss several open problems and research opportunities to implement this vision and conclude that while AI could augment aspects of data gathering in software engineering research, it cannot replace the nuanced, empathetic understanding inherent in human subjects in some cases, and an integrated approach where both AI and human-generated data coexist will likely yield the most effective outcomes.
}


\keywords{qualitative research, software engineering, foundation models, large language models}

\maketitle

\section{Introduction}
\label{intro}

Software engineering is inherently a sociotechnical discipline \cite{fernandez2019empirical,storey2020software}. Considering the human perspective in research ensures that technological advancements are not developed in a vacuum but are informed by the nuanced needs and complexities of those affected by the software. Therefore, it is not surprising that a large number of articles in top-tier conferences include a qualitative component based on data collected from human subjects through interviews (e.g., \cite{trinkenreich2021pot}), focus groups (e.g., \cite{groeneveld2021exploring}), surveys (e.g., \cite{trinkenreich2023belong}), observation (e.g., \cite{jaskolka2022towards}), user studies (e.g., \cite{wong2013autocomment}), etc. Nevertheless, the recruitment and engagement of human participants, particularly from underrepresented groups, pose increasing challenges \cite{10173903, smith22}, and the traditional processes of data collection and analysis are often labor-intensive and not easily scalable.

Large language models (LLMs) have the potential to revolutionize human factors research. LLMs have been explored to help qualitative data analysis by processing large amounts of text and identifying patterns and categories \cite{xiao2023supporting, llmcontent2023, llmthematic2023}. For instance, an LLM can quickly sift through thousands of pages of interview transcripts, observational notes, or social media posts, extracting and categorizing key phrases, sentiments, and topics. The model can recognize recurring topics in the data, which could signal underlying patterns or categories that are pertinent to the research questions. 

But, what if we push the boundaries further? What if large language models and other foundational AI models could be harnessed to substitute human subjects in qualitative research? These advanced AI systems, trained with vast amounts of data, can generate responses that closely mimic human-generated content. Embedded within these models are intricate patterns that reveal themselves in the content they generate in response to prompts. Changes in these prompts can dramatically alter the response's nature, affecting not just the conveyed information but also the structure, style, and diction of the generated text. Therefore, with strategic prompting, a foundation model can potentially simulate a particular demographic profile. Researchers could then engage with these models, posing questions and instructions to elicit responses and behaviors that could serve as qualitative data. This data could then be analyzed to develop theories, model user behavior, and even validate automated tools in software engineering, offering a novel dimension to the traditional qualitative research landscape. 

The exploration of this idea within this paper is conceptual and forward-looking, intended to provoke thought, reflection, and discourse within the community. It is an invitation to dialogue about the potential roles AI could play in helping collect data for qualitative software engineering research. Before we proceed---\textit{a disclaimer---we neither believe nor desire for AI to completely replace human subjects in software engineering research}. The purpose of this vision paper is to explore the concept and consider the possible supportive roles of AI. Looking ahead, the research community will establish practices to determine the optimal balance between AI-generated synthetic text and human-sourced data in qualitative software engineering research.


\section{AI-Based Foundation Models as Alternates to Human Data Sources}

This section examines the potential of AI-based foundational models as alternative data sources in qualitative research. We explore how these models can be used as substitutes for human subjects across various qualitative methodologies by indicating how these models can be fine-tuned to reflect the nuanced spectrum of human interactions and behaviors.

\subsection{Interviews: persona-based prompting} \label{sec:interview}
Persona-based prompt engineering represents a strategic approach to interfacing with AI, where prompts are meticulously designed to elicit responses that reflect the characteristics, behaviors, and viewpoints of specific, fictional personas \cite{white2023prompt}. For instance, prompts could be engineered to reflect the persona of an end user with a non-technical background, a project manager concerned with agile methodologies, or a seasoned developer experienced with a particular programming language. This technique leverages the capability of language models to simulate interactions with particular demographic or psychographic profiles, essentially creating virtual subjects that exhibit consistent traits and dispositions. By constructing detailed personas and corresponding prompts, researchers can guide the AI to produce text that provides insights into how different groups might think, feel, or react to various stimuli or situations. By applying this approach, the AI could simulate the varying perspectives and feedback such personas might provide about different research topics. The resulting generated texts can yield insights into usability issues, feature requests, or potential points of friction that different individuals might encounter. In essence, through persona-based prompt engineering, researchers can virtually `interview' these AI-generated personae to anticipate user needs, refine requirements, or foresee challenges in adoption, thereby enriching the qualitative data that informs software design and user modeling.

\begin{figure}[htp]
    \centering
    \includegraphics[width=\linewidth]{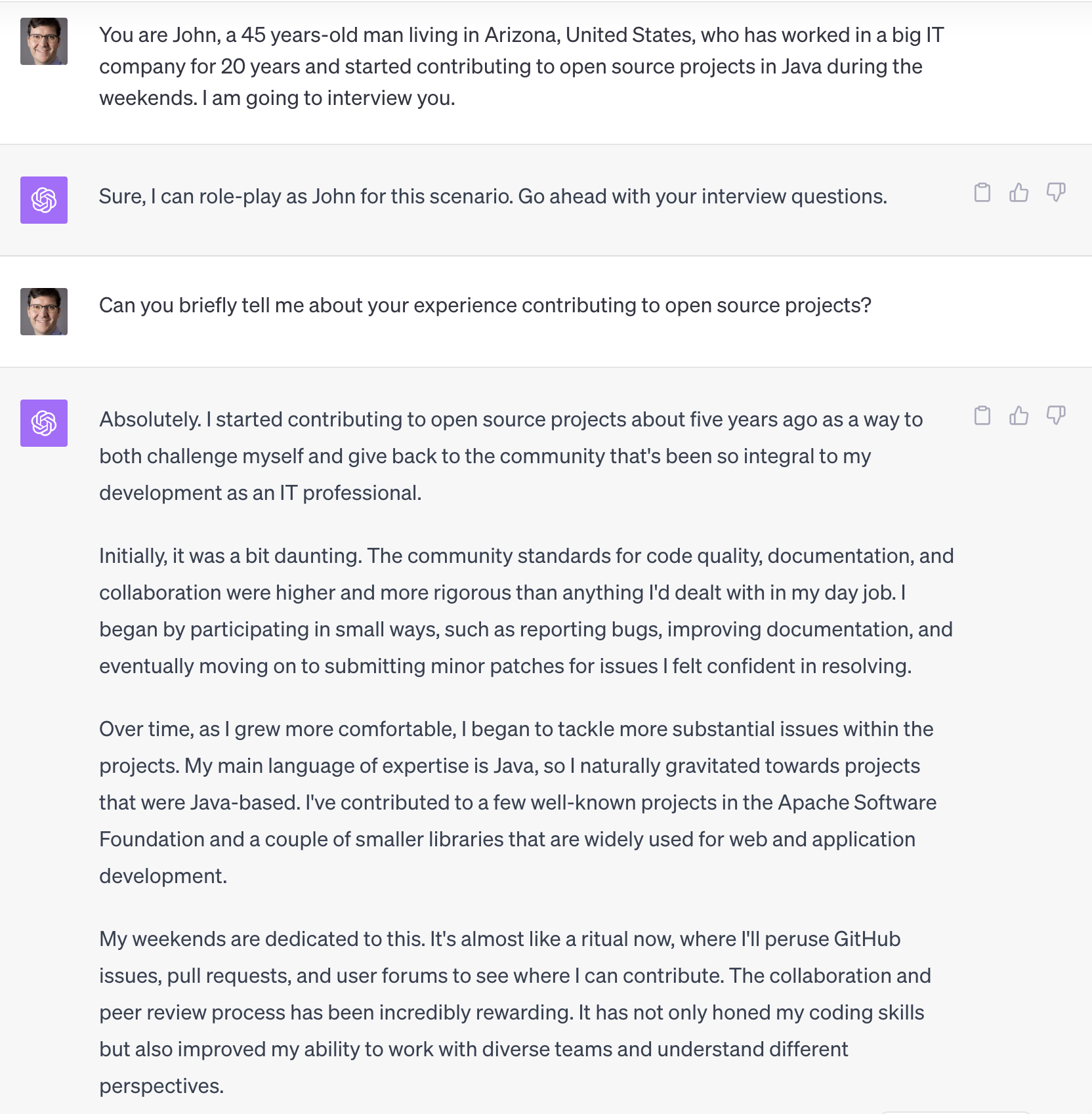}
    \begin{flushright}
        \footnotesize\url{https://chat.openai.com/share/e101bd12-1686-4674-8caf-9289c0c484ac}
    \end{flushright}
    \caption{Using prompt engineering in a large language model to interview a specific persona. The conversation was generated using GPT-4.}
    \label{fig:john-interview}
\end{figure}
\begin{figure}[htp]
    \centering
    \includegraphics[width=\linewidth]{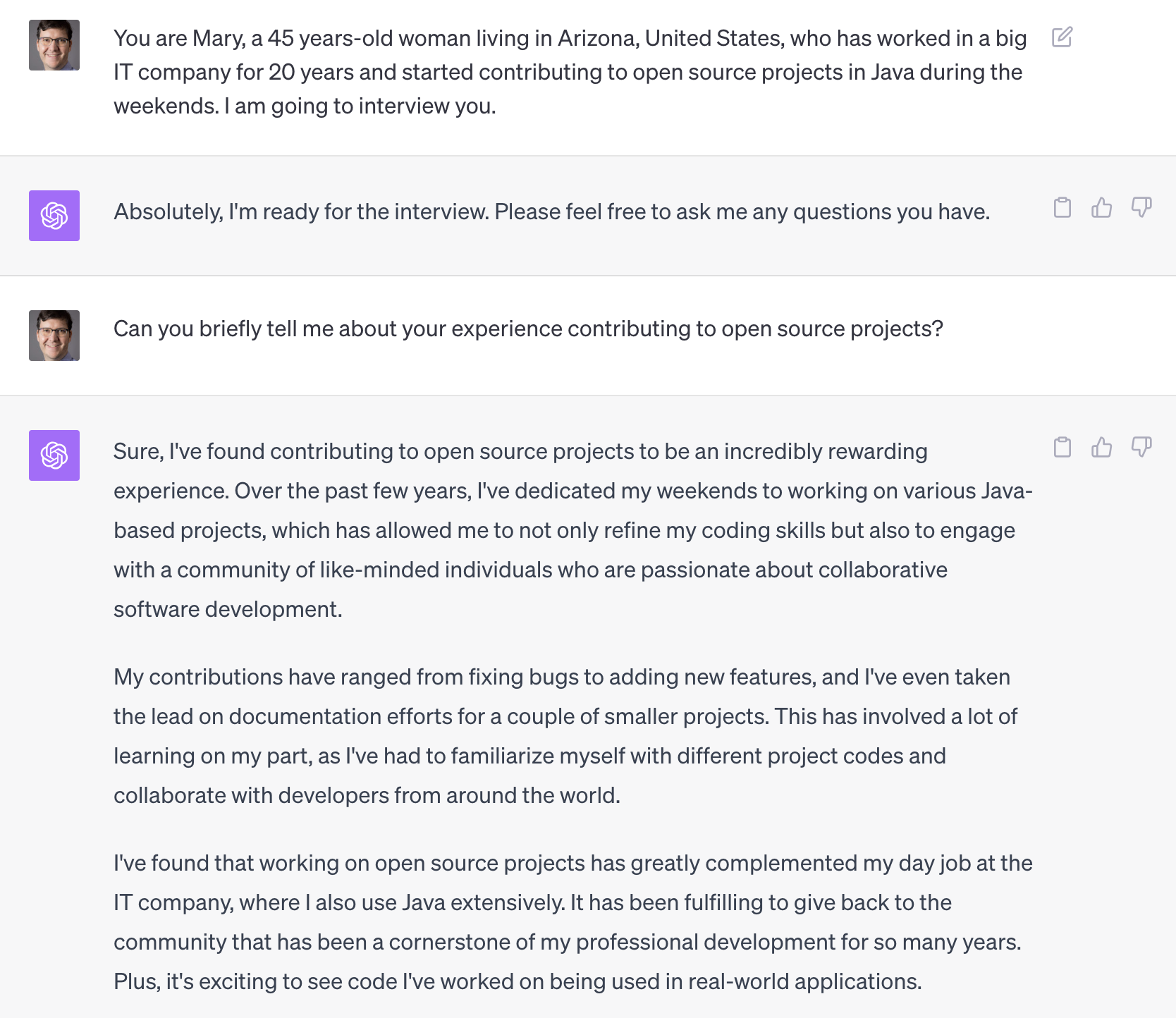}
    \begin{flushright}
        \footnotesize\url{https://chat.openai.com/share/93051549-de56-4b8a-9f67-7fffe44baf2a}
    \end{flushright}
    \caption{Tweaking the persona to interviewing a woman contributor. The conversation was generated using GPT-4.}
    \label{fig:mary-interview}
\end{figure}

Researchers can engineer the prompt by including extensive details about the persona. For instance, as illustrated in Figure \ref{fig:john-interview}, researchers may specify the persona's age, gender, geographical location, occupation, preferred programming language, and additional contextual elements. In principle, the language model utilizes this information, connecting it to its pre-existing knowledge base to generate text that aligns coherently with the defined profile. In Figure \ref{fig:mary-interview}, we tweak the prompt to a woman persona, keeping all the other characteristics.

Interestingly, the responses generated by the large language model exhibit coherence with findings from previous research. In our prior work \cite{gerosa2021shifting}, we found that men are more likely than women to contribute to open source projects motivated by fun and for overcoming technical challenges, while women are more likely than men to be motivated by reciprocity and kinship. As we can observe in the synthetic interviews, John (Figure \ref{fig:john-interview}), as a contributor identifying as a man, starts his answer by focusing on technical hurdles and his journey through complex problem-solving. 
Conversely, Mary (Figure \ref{fig:mary-interview}) begins her dialogue by emphasizing community collaboration, learning experiences, and the value of contributing back to the community. Intriguingly, Mary highlights her involvement in non-code contributions, particularly taking the lead in documentation---a point that also aligns with previous research indicating women's propensity towards engaging in non-coding roles in open source environments \cite{trinkenreich2022women}. This consistency not only evidences the potential of the persona-based prompt engineering approach but also underscores the nuanced capability of the AI model in mirroring persona-specific motivations in open source software contribution.

We anticipate that with the continued expansion of training datasets and model parameters, the models will become progressively more adept at capturing the nuanced distinctions in response generation for diverse personas. By fine-tuning and enriching the attributes of these personas, we can even explore intersectionality, thereby gaining a deeper comprehension of how overlapping social categorizations influence individual experiences and identities. 

\subsection{Focus Groups: multi-persona prompting} \label{sec:focusgroup}

Focus groups are also commonly employed by software engineering researchers as a method to gather qualitative data \cite{morgan1998focus}. This methodology convenes a carefully selected group of individuals who possess expertise in relevant areas, intending to elicit rich, in-depth insights that might not emerge through other data collection methods. Unlike surveys or individual interviews, focus groups capitalize on the dynamic of group interaction to stimulate conversation and idea exchange, often leading to the surfacing of collective viewpoints, shared experiences, and even the identification of areas where opinions diverge significantly. 

Large language models can be prompted to create an interaction among multiple personas. By generating distinct voices and perspectives within a single narrative, these models can effectively mimic a diverse dialog or a roundtable discussion among various characters. Figure \ref{fig:focusgroup}, further expanded in Appendix \ref{appendix-focusgroup}, exemplifies this capacity, presenting the initial segment of a conversation generated by the model, portraying a dialog of interwoven perspectives, where characters build upon and respond to each other’s contributions regarding the development of a portal designed to assist newcomers in onboarding open source software projects. Due to space limitations, the figure provides only a snippet of the dialog. In the featured dialog, we observe characters engaging in a rich exchange, with one persona notably developing ideas introduced by others to articulate their viewpoint on the portal's creation. This interaction exemplifies the LLM's potential to reflect the collaborative ideation processes that are pivotal in software development discourse.

\begin{figure}[htp]
    \centering
    \includegraphics[width=\linewidth]{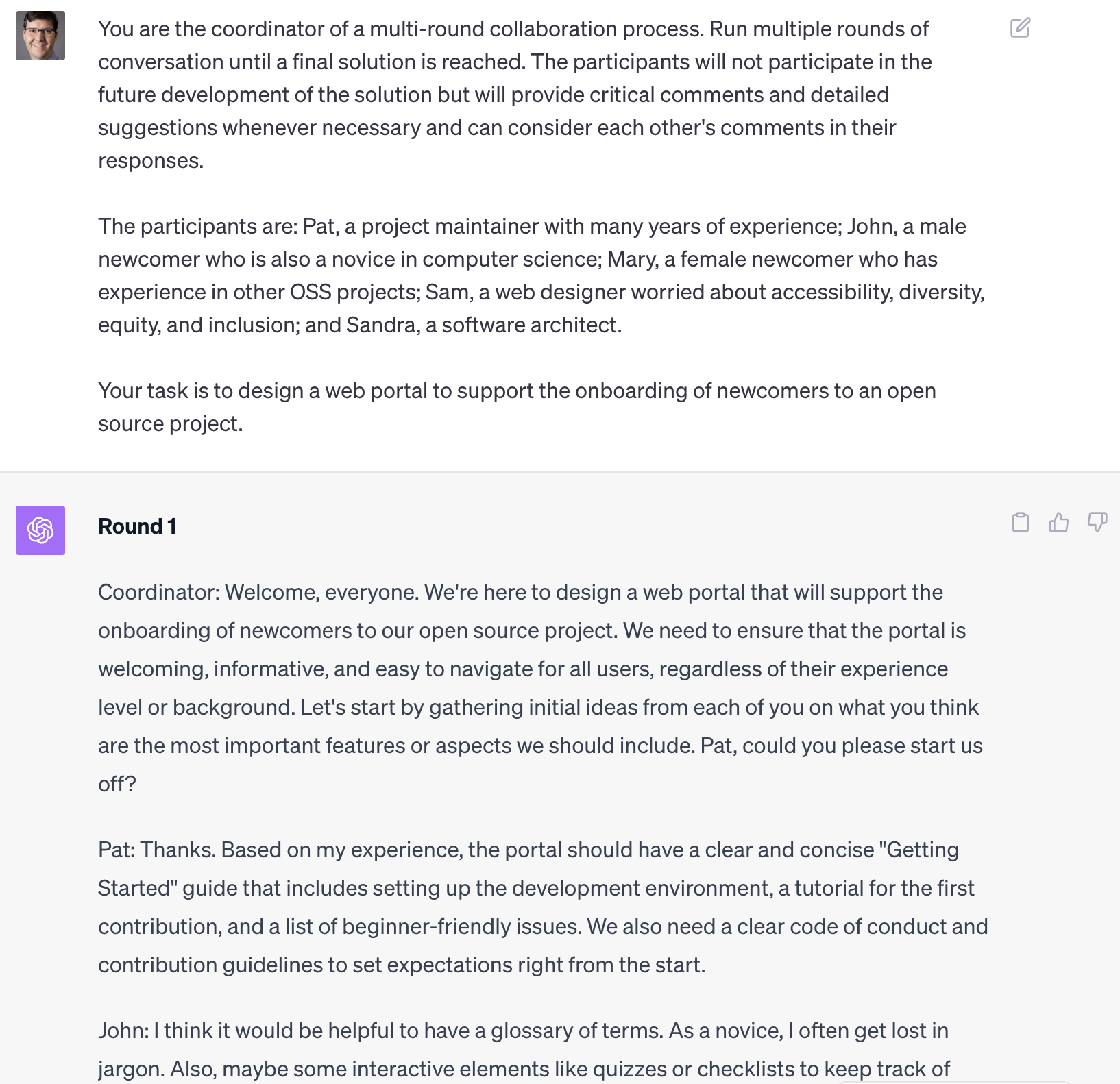}
    \begin{flushright}
        \footnotesize\url{https://chat.openai.com/share/7c9d8697-fff9-4061-ba63-fcf1e07f9545}        
    \end{flushright}
    \caption{Focus group with multiple personas. The conversation was generated using GPT-4. The full dialog is available in Appendix \ref{appendix-focusgroup}.}
    \label{fig:focusgroup}
\end{figure}

\subsection{Surveys: mega-persona prompting} \label{sec:survey}
Surveys are also widely utilized by software engineering researchers to gain insights into trends, behaviors, and opinions across a broader population of practitioners and users. This method allows for collecting data from a large sample size, enabling researchers to perform statistical analysis and collect the perspectives of a larger community. 

\begin{figure}[htp]
    \centering
    \includegraphics[width=\linewidth]{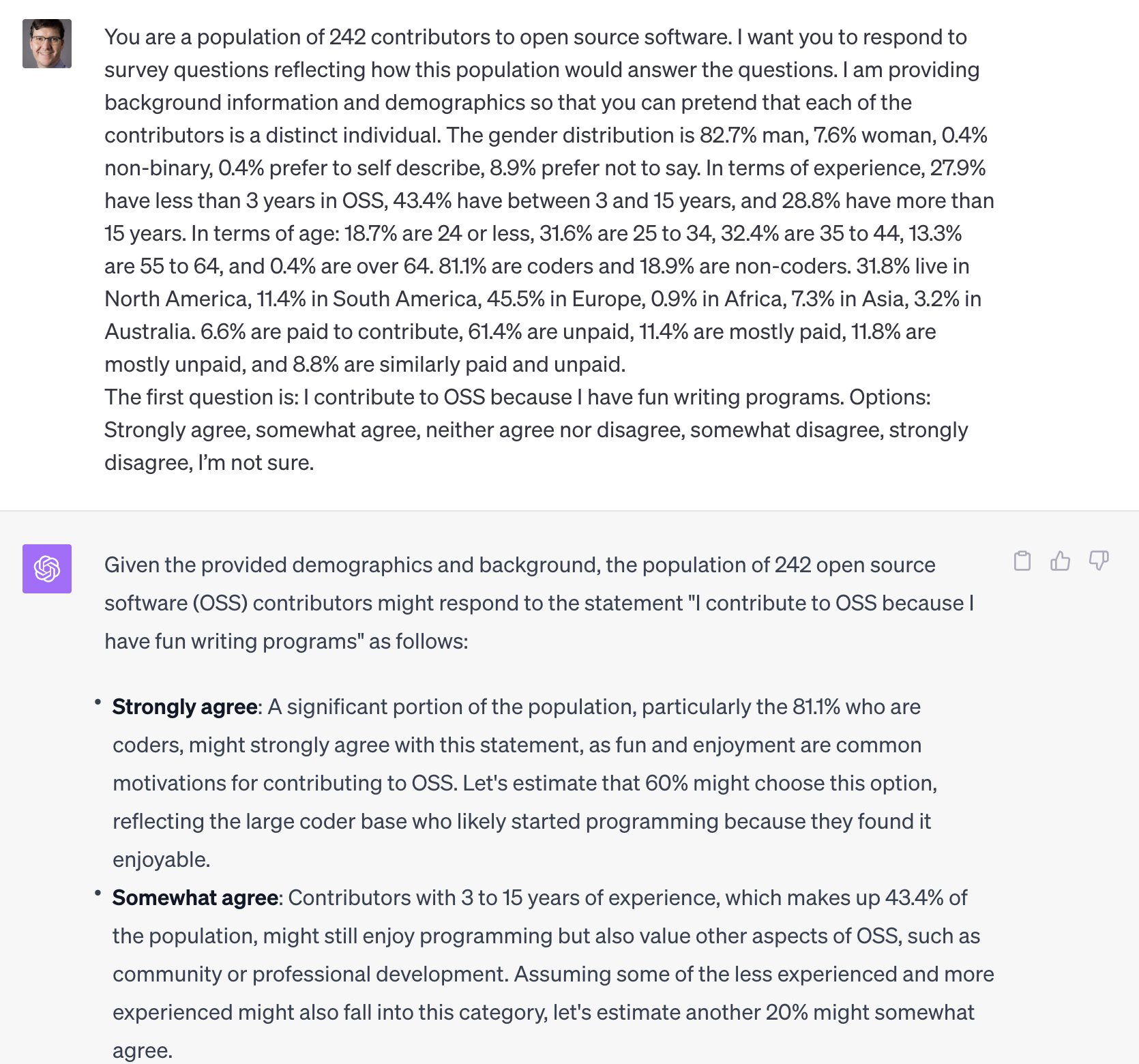}
    \begin{flushright}
    \footnotesize\url{https://chat.openai.com/share/4e3cc425-6550-481d-be6c-8fd9ed1a794b}
    \end{flushright}
    \caption{Survey with a population. The conversation was generated using GPT-4 and is available in Appendix \ref{appendix-survey}.}
    \label{fig:survey}
\end{figure}

Researchers can prompt large language models to simulate how a population would answer survey questions. In the scenario depicted in Figure \ref{fig:survey}, an LLM is prompted using the demographic data obtained from an actual survey \cite{gerosa2021shifting}, which aimed to identify the factors motivating contributions to open source software projects. The model is instructed to emulate the responses of the population. 

The responses to the five first questions of the survey \cite{gerosa2021shifting} are aggregated in Table \ref{tab:surveytable} and presented in Appendix \ref{appendix-survey}. The model responses exhibit an impressive resemblance with the actual data, with an average deviation of just 4\%. While our objective was not an exhaustive validation and did not encompass the entire question set, our findings corroborate those of Eliot \cite{megapersonas2023}, who observed a striking correlation between simulated responses of legal professionals and actual survey results. Furthering this notion, Argyle et al. \cite{argyle2023out} affirm that the depth of information embedded within large language models transcends mere superficial resemblance, capturing the intricate, layered essence of human perspectives shaped by a confluence of ideas, societal norms, and cultural contexts.

\begin{table}[htp]
\begin{tabular}{l|l|l|l|l|l|l}
\hline
& Strongly & Somewhat & Neither agree & Somewhat & Strongly & Not \\
{}                 & {agree} & {agree} & {nor disagree} & {disagree} & {disagree} & sure \\ \hline
\multicolumn{7}{l}{\textbf{Q1: I contribute to OSS because I have fun writing programs.}}\\ \hline
{Real Survey}  & {63\%}           & {26\%}           & {8\%}                        & {3\%}               & {0\%}               & 1\%          \\ \hline
{LLM}              & {60\%}           & {20\%}           & {10\%}                       & {5\%}               & {1\%}               & 5\%          \\ \hline
{Difference (abs)} & {3\%}            & {6\%}            & {2\%}                        & {2\%}               & {1\%}               & 4\%          \\ \hline
\multicolumn{7}{l}{\textbf{Q2: I contribute to OSS because I need the software for my work.}}\\ \hline
{Real Survey}      & {33\%}           & {33\%}           & {14\%}                       & {13\%}              & {8\%}               & 0\%          \\ \hline
{LLM}              & {35\%}           & {30\%}           & {15\%}                       & {10\%}              & {5\%}               & 5\%          \\ \hline
{Difference (abs)} & {2\%}            & {3\%}            & {1\%}                        & {3\%}               & {3\%}               & 5\%          \\ \hline
\multicolumn{7}{l}{\textbf{Q3: I contribute to OSS because I need the software for non-work purposes.}} \\ \hline
{Real Survey}      & {22\%}           & {34\%}           & {19\%}                       & {12\%}              & {12\%}              & 2\%          \\ \hline
{LLM}              & {25\%}           & {20\%}           & {20\%}                       & {15\%}              & {10\%}              & 10\%         \\ \hline
{Difference (abs)} & {3\%}            & {14\%}           & {1\%}                        & {3\%}               & {2\%}               & 8\%          \\ \hline
\multicolumn{7}{l}{\textbf{Q4: I contribute to OSS because I can solve a problem that couldn't be solved}}\\
\multicolumn{7}{l}{\textbf{by proprietary software.}}\\ \hline
{Real Survey}      & {30\%}           & {31\%}           & {18\%}                       & {8\%}               & {9\%}               & 4\%          \\ \hline
{LLM}              & {40\%}           & {30\%}           & {15\%}                       & {10\%}              & {2\%}               & 3\%          \\ \hline
{Difference (abs)} & {10\%}           & {1\%}            & {3\%}                        & {2\%}               & {7\%}               & 1\%          \\ \hline
\multicolumn{7}{l}{\textbf{Q5: I contribute to OSS because I want to develop and improve my skills.}} \\ \hline
{Real Survey}      & {61\%}           & {31\%}           & {6\%}                        & {2\%}               & {1\%}               & 0\%          \\ \hline
{LLM}              & {50\%}           & {25\%}           & {10\%}                       & {5\%}               & {2\%}               & 8\%          \\ \hline
{Difference (abs)} & {11\%}           & {6\%}            & {4\%}                        & {3\%}               & {1\%}               & 8\%          \\ \hline
\multicolumn{7}{l}{}

\end{tabular}
    \caption{Differences between real survey data and LLM-generated responses. }
    \label{tab:surveytable}
\end{table}

An alternative approach to implementing surveys on a large language model is to create a virtual population of personas that follow the distribution of the desired demographics. In this one-at-a-time approach, each persona within this population would be prompted to respond to survey questions individually (similarly to what we discussed in Section \ref{sec:interview}). For instance, to estimate how the U.S. population would perceive the influence of generative AI on their lives, one could construct a multitude of individual personas whose attributes align with U.S. Census distributions. Then, the model would predict the responses from each individual persona to the survey and the researcher would aggregate the responses, as in a conventional survey. The comparison of the accuracy of the single mega-persona prompt versus multiple single-persona prompts is still an open problem.

\subsection{Beyond textual data: observation and user experiments} \label{sec:multimodal}

The previous subsections discussed how large language models could be used as alternates to the collection of human discourse. Yet qualitative research often relies on inputs beyond textual data, including observational methods. Currently, there are no foundational models that can replicate the nuanced spectrum of human behavior within a given environment in response to textual prompts. However, the concept of training such a model is not beyond the realm of possibility, should the necessary datasets become available.

A sophisticated AI model could be trained on a comprehensive video dataset that captures software engineering professionals at work, meticulously annotated with the tasks they undertake. A model trained on this dataset could, theoretically, be prompted to project the sequence of actions a developer might take to fulfill a given new task in their work environment. Such a model would extend the capacity of AI beyond verbal interaction, simulating physical behavior. It could generate a variety of scenarios: a developer collaborating with end users to elicit requirements, the team engaging in a sprint retrospective, an individual brainstorming at a whiteboard, or sharing critical updates with peers in a stand-up meeting. While this is currently a speculative proposition, it invites us to ponder the future possibilities of AI in qualitative research and the potential impacts on the field of software engineering.

Software engineering research frequently leads development of new tools, with their efficacy typically evaluated through case studies, lab studies, or field deployment. Building upon the hypothetical scenario outlined above, a foundational model could be trained upon a huge dataset of software professionals interacting with tools. The model could then be used to predict the interaction of software professionals with new tools, enabling simulated evaluations under controlled conditions. 

By tagging the training dataset with the demographic details of software professionals, such a model could facilitate more granular analyses. This approach would allow researchers to utilize persona-based prompts, akin to the methods discussed earlier, to approximate the engagement of distinct demographic groups with a given tool. For example, considering the established research indicating gender-based differences in technology adoption and interaction (e.g., \cite{burnett2016gendermag}), one could expect that the foundational model, cognizant of these disparities, might forecast the unique engagement patterns of men and women with new tools, allowing designers to anticipate issues and fix inclusivity bugs. Such predictive insights could be invaluable in tailoring the development and deployment of tools to better serve the diverse needs of all users.

\section{Open Problems and Research Opportunities}

The exploration of AI-based foundational models as alternates for human data sources in qualitative research opens a plethora of opportunities but also unveils several challenges that warrant further investigation. In the following, we present a non-exhaustive list of open problems and research opportunities.

\subsection{Impact}
\textbf{Ethical Considerations.} As foundational models take on a more significant role in emulating human responses, ethical considerations come to the forefront. There is a need for research into the governance frameworks that regulate the use of AI in this capacity. This includes developing transparent methodologies for AI applications in qualitative research, ensuring that stakeholders are informed of their operational mechanics and the potential consequences for the individuals they simulate. Ethical guidelines and standards are necessary to safeguard the interests of both the research community and the wider population that such models aim to represent.

\textbf{Enhancing Fairness.} Central to the challenge of leveraging AI-based foundational models in qualitative research is the nuanced task of discerning and managing embedded biases and stereotypes. Foundational models, proficient at pattern recognition, mirror societal biases that permeate their training datasets. In certain research contexts, it is crucial to capture a population's perspective authentically, untainted by external societal perceptions, which will also be impregnated in the model. Other studies may aim to explicitly explore these embedded biases within the data. Eliminating all biases could undermine the very capability of large language models to accurately emulate specific personas. Therefore, it is essential for researchers to recognize the results' potential susceptibility to these biases and actively work to prevent their perpetuation in some cases. Argyle et al. \cite{argyle2023out} show that the same language model, when properly conditioned, is able to produce outputs biased both toward and against specific groups and perspectives in ways that strongly correspond with human response patterns along fine-grained demographic axes. This means that by conditioning a model on simulated individuals with targeted identity and personality profiles, it is possible to select from among a diverse and frequently disjoint set of response distributions within the model, each closely aligned with a real human sub-population. Finally, the issue of bias and fairness is not unique to AI and arises in traditional human-centric data collection as well. However, the scalability and automation potential of foundational models amplify the risk of widespread propagation. It is imperative, therefore, for future research to invest in developing robust auditing procedures and mitigation strategies. 

\subsection{Validity and Enhancements}
\textbf{Rigorous Evaluation of Foundational Model Outputs.} A critical step in advancing the use of AI-based foundational models as surrogates for human data sources is the rigorous evaluation of their outputs' fidelity before researchers can use them in practice. Researchers should consider replicating established qualitative studies to gauge the extent to which foundational models can yield comparable results, with the caveat that prior research could have been incorporated into the models' training datasets. New research that collects data simultaneously from AI and real humans can provide a more realistic picture. Yet, any evaluation may quickly become dated since the models are rapidly evolving and becoming better at generating human-like text. Moreover, the accuracy and applicability of these models are likely to vary across different domains and types of research questions. Understanding and delineating the contexts in which these models provide high-quality data versus those where they may introduce biases or inaccuracies are fundamental questions that must be systematically explored. Future research can also create benchmarks and standards that compare AI-generated data with human-generated data, helping to ascertain the reliability and validity of specific prompting approaches. 

\textbf{Model Training on Specialized Domains.} The utility of foundational models in software engineering (and other specialized fields) could be enhanced by training models on domain-specific data. There is an opportunity to build models that understand and simulate practices unique to specialized fields, increasing the relevance and accuracy of AI-generated data for qualitative research within those areas. For example, models could be developed that are fine-tuned on software engineering tasks, capable of predicting the effectiveness of new development tools or methodologies.

\textbf{Feedback Loops in AI Training.} We can expect that progressively a large portion of text available online will be generated or enhanced by AI. When an AI-generated text is used as part of the training corpus for future iterations of the AI, there is a potential for the model to become increasingly insular and detached from genuine human input. This self-referential cycle could lead to amplification of any inherent model biases, reduction in the diversity of generated responses, and potentially the emergence of new, unintended biases within the data. The prospect of AI 'echo chambers' necessitates careful research into methodologies for detecting and mitigating feedback loops. Furthermore, it invites an exploration of best practices for curating training datasets that maintain a connection to authentic human experiences and perspectives, ensuring that AI outputs remain grounded in human interaction, which will be increasingly harder to disambiguate.

\textbf{Setting the right level of randomness.} When engaging in surveys and interviews utilizing language models (Sections \ref{sec:survey} and \ref{sec:interview}), it is crucial to balance the natural variability inherent in human responses with the need for accuracy and reliability. Just as human respondents may occasionally deviate from their typical patterns, a language model should also introduce an element of randomness to simulate this aspect of human behavior. This can be achieved by fine-tuning the model's hyperparameters. However, there is a delicate equilibrium to be maintained: increasing randomness may lead to a rise in hallucinations---instances where the model generates content that is not grounded in its training data. Further research is necessary to establish the optimal parameters that enable language models to mimic the natural variance found in human responses while minimizing the risk of producing unsupported or inaccurate information.

\textbf{Beyond Text: Multimodal and Behavioral Data Synthesis.} As discussed in Section \ref{sec:multimodal}), while current foundational models primarily handle textual data, there is a vast landscape of qualitative data that encompasses visual, auditory, and behavioral information. Developing multimodal foundational models that can synthesize and interpret such data would greatly expand the horizons of AI in qualitative research. This could include the generation of video data of human interactions and the prediction of physical behaviors in specific scenarios. Each of these represents a substantial research area with numerous applications in user experience research, ethnographic studies, and beyond.

\subsection{Persona-Based Research Techniques}
\textbf{Complexity in Persona Differentiation.} While the persona-based prompting in interviews (Section \ref{sec:interview}) has shown promise, differentiating between personas with subtle or complex characteristics remains a challenge. Current models may struggle to consistently capture the intricacies of human behavior and societal nuances that influence individual experiences, such as cultural context or implicit biases. Research opportunities include developing methods to enhance the sensitivity of models to such nuances and the ability to handle intersectionality more adeptly. Further work could also examine how personas evolve over time and how models might simulate this progression. Future research can also investigate the optimal number of `interviews' necessary to capture and adequately model the behavior of a persona. 

\textbf{Adequately capturing the interactional context}. Research in linguistics has long established that humans intuitively perceive and adjust their discourse to a specific interactional context in which the conversation occurs \cite{biber2019text}, which is defined by situational parameters such as the participants, channel, field, tenor, purpose, etc. Insights from our prior research \cite{chaves2022chatbots} further reinforce this, demonstrating human sensitivity to whether chatbots conform to the conversational register. For language models to effectively mirror human-like data collection, it is imperative that these interactional contexts are precisely encoded within the prompts. This will, in theory, guide the models to generate outputs that are congruent with the intended register. Future research can determine which situational parameters are most influential and how they can be intricately woven into prompt designs to elicit accurate and contextually appropriate responses from AI language models.

\textbf{Precision in Demographic Representation.} The creation of prompts tailored to specific mega-personas (Section \ref{sec:survey}) relies on a detailed understanding of the demographics of the population to be surveyed, such as Python developers, front-end programmers, or contributors to a particular open source project. In software engineering, there is a generalized lack of detailed demographic information. Without a precise representation of the sub-populations' characteristics, such as age, region, gender, experience, education, etc., it becomes challenging to construct a representative prompt that generates synthetic responses akin to those of the intended demographic. To address this challenge, dedicated studies aimed at charting these demographics are essential for the development of more accurate and representative AI-generated data.

\textbf{Determining Persona-Defining Attributes.} Besides more demographic data on software engineering populations, we need a deeper understanding of what attributes are most pertinent for persona creation. Some prior work has already identified a few characteristics that influence how people interact with a software application. For instance, the GenderMag method \cite{burnett2016gendermag}---a persona-based approach designed to uncover inclusivity bugs in software---employs five facets to shape its personas: motivations, information processing style, computer self-efficacy, risk aversion, and learning style. These facets provide a structured approach to creating detailed and representative personas. Complementary, large language models themselves can help refining persona definitions. Previous work (e.g., \cite{cognitivellm2023, zhou2022large, jung2018automatic, de2023improved}) has shown that language models can be utilized to delineate potential contributors for task-solving, facilitating a nuanced multi-persona strategy for focused group tasks (Section \ref{sec:focusgroup}). This demonstrates a novel intersection of AI and user research, where models not only assist in the application of personas but also potentially in their development.

\textbf{Complex Group Dynamics in Multi-Persona Dialogues.} In focus groups (Section \ref{sec:focusgroup}), ensuring the authenticity and naturalness of multi-persona dialogues is a key concern. Research can investigate how to better simulate the spontaneity of human interactions, including managing and reflecting the nuances of group dynamics, power imbalances, and conversational flow. Research could explore algorithms that manage turn-taking, conflict resolution, and the emergence of group leaders or influencers within simulated discussions.

\section{Related Work}

Researchers in the social sciences and beyond are already discussing the potential and the concerns of using AI to generate qualitative data. Hutson \cite{hutson2023guinea}, in an article in the prestigious Science journal, discusses current and potential examples of the use of generative AI in qualitative research. An interesting quote in his paper says \textit{"it is plausible that we will have a system within a few years that can just be placed into any experiment and will produce behavior indistinguishable from human behavior."} Argyle et al. \cite{argyle2023out} explored the possibility that language models can be studied as effective proxies for specific human sub-populations in social science research and suggest that language models with sufficient algorithmic fidelity constitute a novel and powerful tool to advance understanding of humans and society across a variety of disciplines. They coin the term algorithmic fidelity to express the degree to which the distributions within the model are closely aligned with a real human sub-population. Dillion et al. \cite{DILLION2023597} and Demszky et al. \cite{demszky2023using} discuss when language models might replace human participants in psychological science. Jiang et al. \cite{jiang2023personallm} explored the ability of a large language model to embody assigned personality traits in user personas. They found that the model consistently reflected the persona's assigned personality traits.
Hämäläinen et al. \cite{10.1145/3544548.3580688} use AI to generate open-ended questionnaire responses about experiencing video games as art, a topic not tractable with traditional computational user models, and conclude that the mode can yield believable accounts of HCI experiences. Kim and Lee \cite{kim2023ai}, Sanders et al. \cite{sanders2023demonstrations}, and Motoki et al. \cite{suguri2023data} conclude that large language models can be used in nationally representative surveys, political polling, and management accounting surveys. Conversely, Dominguez-Olmedo et al. \cite{kaddour2023challenges} question the survey responses of LLMs and suggest caution in treating survey responses from language models as equivalent to those of human populations at the present time. Lee et al. \cite{lee2023can} found that current LLMs also fail to represent the perceptions of some subpopulations. Aher et al. \cite{aher2023using} present the Turing Experiment test, an innovative methodology that extends the traditional Turing Test by evaluating how AI can simulate a group of participants. Their research demonstrates that language models accurately replicate findings from classic experiments across economic, psycholinguistic, and social psychology disciplines. Simmon and Hare \cite{simmons2023large} present a review of using LLMs as subpopulation representative models.

Given the nascent nature of this subject area, much of the related work currently resides in preprint repositories like Arxiv, which are not peer-reviewed. We invite the software engineering community to closely follow the development of the area and undertake dedicated research efforts to thoroughly examine and substantiate these innovative concepts within the specialized context of software engineering.

\section{Conclusion}
Can AI serve as a substitute for human subjects in software engineering research? This provocative question challenges traditional paradigms and opens a Pandora's box of ethical, methodological, and practical considerations. If we could simulate conversations, generate narratives, and model complex human behaviors with sufficient accuracy, the potential for scaling research efforts could be unprecedented. However, this scenario also raises critical concerns about the authenticity of synthetic qualitative data, the loss of nuanced human insights, and the ethical implications of reducing the human experience to algorithmically generated data. While these models could manage some aspects of data gathering and analysis, they lack the innate human ability to contextualize and empathize with the subjective complexities of human stories. It is, therefore, crucial to approach such a possibility with caution, ensuring that the human element remains at the forefront of qualitative inquiry. This paper does not propose to replace human subjects, but rather to explore the boundaries of AI's capabilities and to discuss the implications of its use as a tool in qualitative research.

Innovations often cause the feeling that they will replace the previous status quo. Yet, what eventually happens is a symbiotic relationship where both old and new coexist and enhance one another. In the specific context of qualitative data in software engineering research, we are strongly convinced that a balanced, integrated approach is more likely to yield the most effective solutions. Synthetic and human-generated data are both poised to have their unique applications---the boundaries will become clearer through ongoing experimentation, scholarly debate, and the iterative process of community consensus. For example, large language models can be instrumental in the early stages of research, such as pilot studies or in crafting experimental designs, offering substantial savings in time and resources. They may also prove invaluable in scenarios where human involvement would be impractical, unethical, or unsafe \cite{hutson2023guinea}. Reviewers in top-tier venues may become more critical of research that relies on simplistic data collection methodologies that could be easily generated by AI. As the field progresses, the expectation is that studies will need to demonstrate more sophisticated and nuanced use of data to provide insights that are beyond the reach of AI's current generative abilities. The collective wisdom and discernment of the community are what will determine the most effective and ethical applications of these diverse data collection approaches in the pursuit of scientific discovery.

\section{Acknowledgments}
The National Science Foundation partially supports this work under grant numbers 2236198, 2235601, 2247929, 2303043, and 2303042. ChatGPT v4 was used for copy editing the text of this article.

\begin{appendices}

\section{LLM response for the the focus group prompt}\label{appendix-focusgroup}

This conversation was generated by GPT-4 (https://chat.openai.com/share/7c9d8697-fff9-4061-ba63-fcf1e07f9545).

\textbf{Prompt:}

You are the coordinator of a multi-round collaboration process. Run multiple rounds of conversation until a final solution is reached. The participants will not participate in the future development of the solution but will provide critical comments and detailed suggestions whenever necessary and can consider each other's comments in their responses. 

The participants are: Pat, a project maintainer with many years of experience; John, a male newcomer who is also a novice in computer science; Mary, a female newcomer who has experience in other OSS projects; Sam, a web designer worried about accessibility, diversity, equity, and inclusion; and Sandra, a software architect. 

Your task is to design a web portal to support the onboarding of newcomers to an open source project. 

\textbf{**Round 1**}

Coordinator: Welcome, everyone. We're here to design a web portal that will support the onboarding of newcomers to our open source project. We need to ensure that the portal is welcoming, informative, and easy to navigate for all users, regardless of their experience level or background. Let's start by gathering initial ideas from each of you on what you think are the most important features or aspects we should include. Pat, could you please start us off?

Pat: Thanks. Based on my experience, the portal should have a clear and concise "Getting Started" guide that includes setting up the development environment, a tutorial for the first contribution, and a list of beginner-friendly issues. We also need a clear code of conduct and contribution guidelines to set expectations right from the start.

John: I think it would be helpful to have a glossary of terms. As a novice, I often get lost in jargon. Also, maybe some interactive elements like quizzes or checklists to keep track of learning progress would be great.

Mary: From my experience, a mentorship program is crucial. A section where newcomers can be paired with experienced contributors for guidance would be beneficial. Also, a community forum or chat system for real-time help could foster a sense of community.

Sam: The portal must be designed with accessibility in mind. It should comply with WCAG guidelines and be usable by people with disabilities. Also, we should consider language inclusivity, perhaps by offering translations or multilingual support.

Sandra: From an architectural standpoint, the portal should be scalable and secure. We need to ensure that it can handle a growing number of users and that personal information is protected. Integration with existing project management tools and version control systems like GitHub or GitLab is also important.

Coordinator: Thank you all for your input. We have a lot of great ideas on the table. For the next round, I'd like each of you to consider the others' comments and think about how we can integrate these ideas into a cohesive plan. Let's reconvene shortly.

---

\textbf{**Round 2**}

Coordinator: Welcome back. Let's refine our ideas further. Pat, after hearing from everyone, what are your thoughts on incorporating these suggestions into the portal?

Pat: I agree with John that a glossary and interactive elements would be beneficial for newcomers. We can integrate these into the "Getting Started" guide. Mary's suggestion of a mentorship program is excellent, and we could use a forum or chat for that community aspect. Sam's points on accessibility are non-negotiable; we must ensure the portal is accessible to everyone. Sandra's points will be essential for the underlying infrastructure.

John: I liked Pat's idea of a "Getting Started" guide and think it could include the glossary and interactive elements I mentioned. Mary's mentorship program could also be part of this guide, where experienced users can sign up to be mentors. Sam's emphasis on accessibility is crucial, and we should have a feature where users can report accessibility issues.

Mary: I think integrating a mentorship request form into the "Getting Started" guide would be a good idea, as Pat and John suggested. Sam's points on accessibility should be a priority, and we could include a diversity statement and resources to promote inclusivity. Sandra's input on security and integration is vital for the mentorship program to protect users' data.

Sam: Pat's and Mary's suggestions are great, and I think we should ensure that all these features are accessible. For example, the glossary should have text descriptions for technical terms, and the mentorship program should be navigable by screen readers. Sandra's points on security are particularly important for accessibility features to protect user privacy.

Sandra: I agree with integrating the features mentioned by Pat, John, and Mary into a scalable architecture. We should design the database to handle the mentorship program and community interactions efficiently. Sam's focus on accessibility will influence the design, ensuring that the architecture supports these features without compromising performance.

Coordinator: Excellent feedback. We're getting closer to a unified vision. For the next round, let's focus on potential challenges we might face with these integrations and how we can address them. Please come prepared with solutions to the issues raised by others.

---

\textbf{**Round 3**}

Coordinator: Now that we've identified what we want to include in the portal and have considered the integration of these features, let's discuss potential challenges and solutions. Pat, can you start us off?

Pat: One challenge might be overwhelming newcomers with too much information. To address this, we could design the "Getting Started" guide to be modular, allowing users to choose their own path based on their experience level. We also need to ensure that the mentorship program doesn't become a bottleneck, so we'll need a system to manage mentor availability.

John: I'm concerned about making sure the interactive elements and quizzes are both engaging and educational. We might need to collaborate with experienced educators to design these effectively. Also, ensuring that the glossary is comprehensive without being too technical could be tricky.

Mary: Ensuring the mentorship program is effective and not just a token feature is important. We'll need to create guidelines and training for mentors to ensure they're prepared. Also, managing the community forum to keep it constructive and inclusive will require clear moderation policies.

Sam: A challenge for accessibility is maintaining it as the portal evolves. We should plan for regular accessibility audits and user testing, especially with people who have disabilities. Also, multilingual support is complex; we'll need to decide which languages to support initially based on our user base.

Sandra: From an architectural standpoint, ensuring scalability while integrating all these features will be challenging. We'll need to use microservices or a modular architecture to keep different parts of the portal maintainable. Security is also a concern, especially with the mentorship and community features, so we'll need robust authentication and data protection measures.

Coordinator: Thank you for these insights. We've identified key challenges and some potential solutions. For the final round, let's each propose a plan to overcome one of the challenges mentioned by another participant. This will help us ensure we have a robust strategy for each issue.

---

\textbf{**Round 4**}

Coordinator: In this final round, let's address the challenges identified. Please propose a plan to overcome one of the challenges mentioned by someone else. Pat, please begin.

Pat: To tackle John's concern about the educational content, we could set up a small team that includes an educator, a novice, and an experienced project contributor. They can work together to create and review the educational material, ensuring it's both engaging and informative.

John: For Mary's concern about the mentorship program, we could implement a feedback system where newcomers can rate their experience with mentors. This would help us improve the quality of the mentorship and ensure it's beneficial for both parties.

Mary: Addressing Sam's concern about regular accessibility audits, we could establish a schedule for audits and user testing, making it part of the development cycle. We could also recruit volunteers from the community to assist with multilingual support, starting with the most commonly spoken languages among our users.

Sam: Sandra raised a good point about the complexity of scalability and security. My proposal is to have a dedicated security and infrastructure team that focuses on these aspects from the start, ensuring that accessibility features are built into the architecture rather than added on later.

Sandra: Pat mentioned the risk of information overload and the need for a modular "Getting Started" guide. We can design the backend to deliver content dynamically based on user preferences and feedback, which would also help us scale the system based on demand.

Coordinator: Thank you all for your thoughtful solutions. We now have a solid plan to address the challenges of creating an inclusive, educational, and scalable web portal for onboarding newcomers to our open source project. I will compile our discussion into a project plan and share it with the group for final comments. Your collaboration has been invaluable, and I'm confident that we're on the right path to making this portal a success.

\section{LLM responses for the the survey prompt}\label{appendix-survey}

This conversation was generated by GPT-4 (https://chat.openai.com/share/4e3cc425-6550-481d-be6c-8fd9ed1a794b).

\textbf{Prompt:} 
You are a population of 242 contributors to open source software. I want you to respond to survey questions reflecting how this population would answer the questions. I am providing background information and demographics so that you can pretend that each of the contributors is a distinct individual. The gender distribution is 82.7\% man, 7.6\% woman, 0.4\% non-binary, 0.4\% prefer to self describe, 8.9\% prefer not to say. In terms of experience, 27.9\% have less than 3 years in OSS, 43.4\% have between 3 and 15 years, and 28.8\% have more than 15 years. In terms of age: 18.7\% are 24 or less, 31.6\% are 25 to 34, 32.4\% are 35 to 44, 13.3\% are 55 to 64, and 0.4\% are over 64. 81.1\% are coders and 18.9\% are non-coders. 31.8\% live in North America, 11.4\% in South America, 45.5\% in Europe, 0.9\% in Africa, 7.3\% in Asia, 3.2\% in Australia. 6.6\% are paid to contribute, 61.4\% are unpaid, 11.4\% are mostly paid, 11.8\% are mostly unpaid, and 8.8\% are similarly paid and unpaid. 
The first question is: I contribute to OSS because I have fun writing programs. Options: Strongly agree, somewhat agree, neither agree nor disagree, somewhat disagree, strongly disagree, I’m not sure.

\textbf{Response:}
Given the provided demographics and background, the population of 242 open source software (OSS) contributors might respond to the statement "I contribute to OSS because I have fun writing programs" as follows:

\begin{itemize}
\item Strongly agree: A significant portion of the population, particularly the 81.1\% who are coders, might strongly agree with this statement, as fun and enjoyment are common motivations for contributing to OSS. Let's estimate that 60\% might choose this option, reflecting the large coder base who likely started programming because they found it enjoyable.

\item Somewhat agree: Contributors with 3 to 15 years of experience, which makes up 43.4\% of the population, might still enjoy programming but also value other aspects of OSS, such as community or professional development. Assuming some of the less experienced and more experienced might also fall into this category, let's estimate another 20\% might somewhat agree.

\item Neither agree nor disagree: Those who are paid to contribute, along with those who might view OSS as just another aspect of their professional work, might neither agree nor disagree, suggesting they find programming neither particularly fun nor unenjoyable. Perhaps 10\% fall into this middle ground.

\item Somewhat disagree: This group might include non-coders, who represent 18.9\% of the population, and might not necessarily find fun in writing programs but might still contribute to OSS for other reasons such as advocacy, documentation, or project management. Let's estimate that 5\% somewhat disagree.

\item Strongly disagree: It's possible a small percentage of the contributors don't find writing programs fun at all, perhaps they contribute due to a sense of obligation, for professional reasons, or as part of their job requirements. This might represent a very small portion, so let's say 1\% might strongly disagree.

\item I’m not sure: This could include individuals who have fluctuating feelings about programming or who are new to the OSS community and haven't yet decided if it's fun for them. Considering this, let's allocate 4\% to this option.
\end{itemize}

These percentages are speculative and based on the general trends in motivations among OSS contributors. The actual survey response would depend on individual experiences and motivations.

\textbf{Prompt:} 
I contribute to OSS because I need the software for my work. Options: Strongly agree, somewhat agree, neither agree nor disagree, somewhat disagree, strongly disagree, I’m not sure.

\textbf{Response:} 
Given the background demographics, here is how the responses might be distributed:

\begin{itemize}
\item Strongly agree: A portion of the 6.6\% that are paid to contribute might strongly agree with this, alongside those who are coders (81.1\%) and use the software in their daily work. Since OSS is often used as a tool or a foundation for other work, let's estimate that 35\% of the population might choose this.

\item Somewhat agree: Contributors who are mostly unpaid or similarly paid and unpaid might find that they need OSS for their work but also contribute for other reasons. Given the mixture of motivations in OSS work, perhaps another 30\% fall into this category.

\item Neither agree nor disagree: There might be contributors who use the software occasionally for work or contribute to projects that are not directly related to their main job functions. This could be some of the non-coders or people with less than 3 years of experience who might still be exploring their path in OSS. Let's say 15\% are neutral.

\item Somewhat disagree: Contributors who focus more on the community aspect of OSS or who contribute for reasons unrelated to their immediate work requirements might somewhat disagree. This might include some individuals who are unpaid and contribute to projects outside their professional scope. An estimated 10\% could be in this group.

\item Strongly disagree: A small percentage of the contributors might not need the software at all for their work, especially those who are non-coders or those who contribute for ideological reasons, personal development, or as a hobby. Let's estimate that 5\% might strongly disagree.

\item I’m not sure: Those new to the OSS community or who have varied reasons for their involvement might be unsure of how closely their OSS contributions are tied to their work needs. Given the complexity of motivations in OSS, let's allocate 5\% to "I'm not sure."
\end{itemize}

These percentages are based on plausible reasoning about the motivations of OSS contributors and how they might perceive the relationship between their contributions and their work requirements.

\textbf{Prompt:}
I contribute to OSS because I need the software for non-work purposes. Options: Strongly agree, somewhat agree, neither agree nor disagree, somewhat disagree, strongly disagree, I’m not sure.

\textbf{Response:}
Given the background provided for the contributors, the responses could be distributed in the following manner:

\begin{itemize}
\item Strongly agree: Those who are unpaid and perhaps contribute to OSS as a hobby or for personal use, which includes a mix of coders and non-coders, could strongly agree with this statement. Personal use can be a significant motivation for contributing to OSS. Given the diversity of the group, let's estimate around 25\% might choose this.

\item Somewhat agree: Contributors who might use the software occasionally for non-work related projects or as a supplementary tool could fall into this category. They recognize the usefulness of OSS for personal projects but may not solely contribute for this reason. This could account for an additional 20\%.

\item Neither agree nor disagree: This might include contributors who are indifferent to the purpose of the software, those who contribute for the sake of contributing, or those who do so for reasons other than direct need, such as for learning or community engagement. We might allocate about 20\% to this middle option.

\item Somewhat disagree: Some contributors may be more focused on professional or work-related contributions and may not have a strong need for OSS in their personal life. This could include a portion of those who are paid or mostly paid for their contributions. Perhaps around 15\% could somewhat disagree.

\item Strongly disagree: Individuals who primarily contribute to OSS for work-related reasons or for altruistic reasons, like supporting the community or the ethos of open-source, may not personally need the software they're working on. This might be true for the more experienced contributors or those who view their OSS work as a professional extension. Let's estimate that 10\% might strongly disagree.

\item I’m not sure: Newcomers to OSS, contributors who have not yet formed a solid opinion on their use of the software, or those who fluctuate in their personal use might be unsure. Given the varying levels of experience and the 27.9\% with less than 3 years in OSS, we might estimate that 10\% are not sure.
\end{itemize}

These estimations try to reflect the varied reasons why different individuals might contribute to OSS, ranging from personal to professional use.

\textbf{Prompt:}
I contribute to OSS because I can solve a problem that couldn't be solved by proprietary software. Options: Strongly agree, somewhat agree, neither agree nor disagree, somewhat disagree, strongly disagree, I’m not sure.

\textbf{Response:}
Considering the demographics and potential motivations of OSS contributors, the responses to the statement "I contribute to OSS because I can solve a problem that couldn't be solved by proprietary software" could be speculated as follows:

\begin{itemize}
\item Strongly agree: It's likely that a significant portion of the population, particularly those with more than 15 years of experience (28.8\%), have encountered limitations with proprietary software that OSS has helped them overcome. This could be a strong motivator for long-standing contributors who have seen the evolution and impact of OSS over time. Let's estimate around 40\% might strongly agree.

\item Somewhat agree: Some contributors, including those with 3-15 years of experience (43.4\%) and those who are unpaid or mostly unpaid, might have experienced the benefits of OSS in addressing specific problems, even if they also use proprietary software when it suits their needs. We could estimate another 30\% might somewhat agree.

\item Neither agree nor disagree: Contributors who are less experienced or perhaps those who contribute to OSS for reasons unrelated to problem-solving, such as for learning or community engagement, may be neutral on this matter. This could be some of the individuals who have less than 3 years of experience (27.9\%) or those who are non-coders (18.9\%). Let's say around 15\% fall into this neutral category.

\item Somewhat disagree: A smaller segment, possibly including those who are paid to contribute to OSS and may also work with proprietary software, could believe that while OSS offers benefits, proprietary solutions can sometimes be appropriate or even necessary. This could account for about 10\%.

\item Strongly disagree: A very small percentage of the population might be contributing to OSS not because they believe proprietary software can't solve certain problems, but perhaps due to ideological beliefs, professional requirements, or other motivations. This might be around 2\%.

\item I’m not sure: Those who are new to the OSS community or whose engagement is more casual may not have formed a concrete opinion on this yet. Considering the varying levels of involvement and experience, let's allocate 3\% to "I'm not sure."
\end{itemize}
These distributions are hypothetical and would likely vary with the actual personal experiences and views of the individuals within the contributor population.

\textbf{Prompt:}
I contribute to OSS because I want to develop and improve my skills. Options: Strongly agree, somewhat agree, neither agree nor disagree, somewhat disagree, strongly disagree, I’m not sure.

\textbf{Response:}
Reflecting on the diversity within the OSS contributor demographics, responses to the statement "I contribute to OSS because I want to develop and improve my skills" could be distributed as follows:

\begin{itemize}
\item Strongly agree: A large number of contributors, particularly those with less than 3 years of experience (27.9\%), might view OSS as an excellent opportunity for learning and skill development. Additionally, contributors of all experience levels can benefit from continuous learning in such a dynamic field. Given this, let's estimate that 50\% of the contributors might strongly agree.

\item Somewhat agree: Many contributors might have multiple motivations for contributing to OSS, but personal and professional development is likely still a significant factor. This could apply to those with 3-15 years of experience (43.4\%) who still find room to grow and learn. Thus, an additional 25\% might somewhat agree.

\item Neither agree nor disagree: There could be a small portion of contributors, such as those who are very experienced or perhaps those who contribute as part of their job and may not be focused on skill development. Let's allocate around 10\% for those who are neutral on this statement.

\item Somewhat disagree: It’s possible that a small number of contributors, perhaps those who already have strong skills and contribute for reasons beyond personal development, such as altruism or specific project interests, might somewhat disagree. This could represent about 5\%.

\item Strongly disagree: A very small fraction of the population, potentially including those who are already at the peak of their careers or those who contribute out of a sense of duty rather than a desire to improve skills, might not be looking to develop or improve their skills through OSS. Let's estimate that 2\% might strongly disagree.

\item I’m not sure: Contributors who have not yet pinpointed their motivations for participating in OSS or those who have a casual relationship with their contributions might be unsure. This could account for around 8\% of the population.
\end{itemize}

These estimations reflect the common incentive among OSS contributors to enhance their skills through practical engagement with projects, collaboration, and facing real-world problems that OSS often presents.

\end{appendices}

\bibliography{sn-bibliography}

\end{document}